\documentclass[prl,letterpaper,aps,twocolumn]{revtex4}
\usepackage{graphicx}
\usepackage[active]{srcltx}
\usepackage{amsfonts}
\usepackage[latin1]{inputenc}
\graphicspath{{FIGURES/}}

\begin{document}
\title[polyhomeostasis]{Self-organized chaos through polyhomeostatic 
                     optimization}

\author{D. Markovic and Claudius Gros}
\affiliation{Institute for Theoretical Physics, 
             Johann Wolfgang Goethe University, 
            Frankfurt am Main, Germany}

\date{\today}

\begin{abstract}
The goal of polyhomeostatic control is to achieve a
certain target distribution of behaviors, in contrast 
to homeostatic regulation which aims at stabilizing a 
steady-state dynamical state. We consider polyhomeostasis
for individual and networks of firing-rate neurons, 
adapting to achieve target distributions of firing
rates maximizing information entropy. We show that any
finite polyhomeostatic adaption rate destroys all attractors
in Hopfield-like network setups, leading to intermittently
bursting behavior and self-organized chaos. The importance 
of polyhomeostasis to adapting behavior in general is discussed.
\end{abstract}
\pacs{89.75.Hc, 02.10.Ox, 87.23.Ge, 89.75.Fb}
\maketitle


{\it Introduction.}--
Homeostatic regulation plays a central role in
all living, as well as in many technical applications.
Biological parameters, like the blood sugar level,
the heart beating frequency or the average firing
rates of neurons need to be maintained within certain 
ranges in order to guarantee survival. The same holds
in the technical regime for the rotation speed of engines
and the velocity of airplanes, to give a few examples.

Homeostasic control in the brain goes beyond the regulation
of scalar variables like the concentration of proteins
and ions, involving the functional stability of neural 
activity both on the individual as well as on a 
network level~\cite{turrigiano99,davis06,marder06}. We use here the
term `polyhomeostasis' for self-regulating processes aimed
at stabilizing a certain target distribution of dynamical
behaviors. Polyhomeostasis is an important concept 
used hitherto mostly implicitly and not yet well studied
from the viewpoint of dynamical system theory.
Polyhomeostasis is present whenever the goal of the
autonomous control is the stabilization of a
non-trivial distribution of dynamical states,
polyhomeostatic control hence generalizes the concept
of homeostasis.
The behavior of animals on intermediate
time scales, to give an example, may be regarded 
as polyhomeostatic, aiming at optimizing a distribution 
of qualitatively different rewards, like food, 
water and protection; animals are not just trying to
maximize a single scalar reward quantity. A
concept loosely related to polyhomeostasis is
homeokinesis, proposed in the context 
of closed-loop motion learning~\cite{der99}, having the 
aim to stabilize non-trivial but steady-state movements
of animals and robots.

Here we study generic properties of dynamical systems governed
by polyhomeostatic self-regulation using a previously proposed
model~\cite{stemmler99,triesch05a} for regulating the
firing-rate distribution of individual neurons based 
on information-theoretical principles. 
We show that polyhomeostatic regulation, aiming at
stabilizing a specific target distribution of neural activities
gives rise to non-trivial dynamical states when
recurrent interactions are introduced. We find, 
in particular, that the introduction of polyhomeostatic 
control to attractor networks leads to a destruction 
of all attractors resulting for large networks, as a 
function of the average firing rate, in either 
intermittent bursting behavior
or self-organized chaos, with both states being
globally attracting in their respective phase spaces.
 

{\it Firing-rate distributions.}--
We consider a discrete-time, rate 
encoding artificial neuron with 
input $x\in[-\infty,\infty]$, output
$y\in[0,1]$ and a transfer function $g(z)$,
\begin{equation}
y(t+1) \ =\ g\big(a(t) x(t)+b(t)\big)~,
\qquad g(z)\ =\ {1\over {\rm e}^{-z}+1}~.
\label{eq_neuron}
\end{equation}
The gain $a(t)$ and the threshold
$-b(t)/a(t)$ in (\ref{eq_neuron})
are slow variables, their
time evolution being determined
by polyhomeostatic considerations.

Information is encoded in the brain through
the firing states of neurons and it is therefore
plausible to postulate~\cite{stemmler99},
that polyhomeostatic adaption
for the internal parameters $a(t)$ and $b(t)$
leads to a distribution $p(y)$ for the firing rate
striving to encode as much information as
possible given the functional form (\ref{eq_neuron})
of the transfer function $g(z)$.
The normalized exponential distribution
\begin{equation}
p_\lambda(y) \ =\ 
{ \lambda\,{\rm e}^{-\lambda y}\over 1-{\rm e}^{-\lambda}},
\qquad
\mu \ =\ {1\over \lambda}
{{\rm e}^{\lambda }-1-\lambda\over {\rm e}^{\lambda}-1}~,
\label{eq_exponential}
\end{equation}
maximizes the Shannon entropy~\cite{cover06},
viz the information content, on the interval $y\in[0,1]$,
for a given expectation value $\mu$.
A measure for the closeness of the two probability 
distributions $p(y)$ and $p_\lambda(y)$ is given by
the Kullback-Leibler divergence~\cite{cover06}
\begin{equation}
D_\lambda(a,b)\ =\ 
\int p(y)\log\left({p(y)\over p_\lambda(y)}\right) dy~,
\label{eq_Kullback_Leibler}
\end{equation}
which is, through (\ref{eq_neuron}), a function of the 
internal parameters $a$ and $b$. The Kullback-Leibler 
divergence is strictly positive and vanishes only when
the two distributions are identical. By minimizing 
$D_\lambda(a,b)$ with respect to $a$ and $b$
one obtains~\cite{triesch05a} the stochastic gradient
rules 
\begin{equation}
\begin{array}{rcl}
\Delta a &=& \epsilon_a \left({1/a} + x\Delta \tilde b\right) \\
\Delta b &=& \epsilon_b\, \Delta \tilde b, \qquad
\Delta \tilde b= 1-(2+\lambda)y+\lambda y^2
\end{array}
\label{eq_intrinsic_plasticity}
\end{equation}
which have been called `intrinsic plasticities'~\cite{turrigiano99}.
The respective learning rates $\epsilon_a$ and $\epsilon_b$
are assumed to be small, viz the time evolution of the
internal parameters $a$ and $b$ is slow compared to the
evolution of both $x$ and $y$. For any externally 
given time series $x(t)$ for the input, the 
adaption rules (\ref{eq_intrinsic_plasticity})
will lead to a distribution of the output firing rates
$y(t)$ as close as possible, given the specification 
(\ref{eq_neuron}) for the transfer function,
to an exponential with given mean $\mu$.

\begin{figure}[t]
\centerline{
\includegraphics[width=0.35\textwidth]{stability.eps}
\hspace{4ex}
\includegraphics[width=0.09\textwidth]{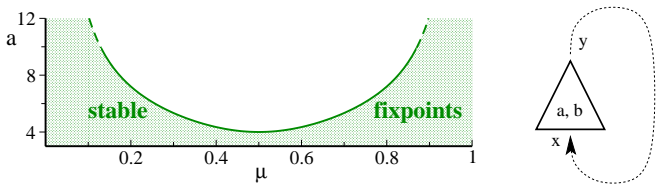}
           }
\caption{(color online) Left: As a function of the average firing
rate $\mu$, the region of stability (shaded area) of the
fixpoint $(y^*, b^*)$, see Eq.\ (\ref{eq_fixpoint_yb}) 
of the one-neuron network in the quasistatic limit.
Right: A self-coupling neuron with internal parameters
$a(t)$ and $b(t)$.
}
\label{fig_stability} 
\end{figure}


{\it Attractor relics.}--
In deriving the stochastic gradient 
rules (\ref{eq_intrinsic_plasticity}) it has been 
assumed that the input $x(t)$ is statistically 
independent of the output $y(t)$ (but not vice versa).
This is not any more the case when a set of polyhomeostatically
adapting neurons are mutually interconnected, forming
a recurrent network, Here we will show that networks
based on polyhomeostatic principles will generically
show spontaneous and continuously ongoing activity.

In a first step we analyze systematically the smallest
possible network, viz the single-site loop, obtained
by feeding the output back to the input,
see Fig.\ \ref{fig_stability}, a setup which has
been termed in a different context `autapse' \cite{seung00}.
We use the balanced substitution $x\to y-1/2$ in
Eqs.\ (\ref{eq_neuron}) and (\ref{eq_intrinsic_plasticity}),
the complete set of evolution rules for 
the dynamical variables $y(t),\, a(t)$ and $b(t)$ 
is then
\begin{eqnarray}
\nonumber
y(t+1) & =& g\big(a(t)[y(t)-1/2]+b(t)\big) \\
\label{eq_one_site}
b(t+1) &=& b(t)+\epsilon_b\,\Delta \tilde b(t) \\
a(t+1) &=& a(t)+\epsilon_a \left({1/a(t)} 
             + [y(t)-1/2]\Delta \tilde b(t)\right) 
\nonumber
\end{eqnarray}
with 
\begin{equation}
\Delta \tilde b(t)\ =\ 1-(2+\lambda)y(t+1)+\lambda y^2(t+1)~.
\label{eq_Delta_tile_b}
\end{equation}
Note, that $\Delta \tilde b(t)$ in (\ref{eq_Delta_tile_b})
depends on $y(t+1)$, and not on $y(t)$, as
one can easily verify when going through the derivation
of the rules (\ref{eq_intrinsic_plasticity}) for the
intrinsic plasticity. The evolution 
equations (\ref{eq_one_site}) are invariant
under $y\leftrightarrow (1-y)$,
$b\leftrightarrow (-b)$,
$a\leftrightarrow a$ and
$\lambda \leftrightarrow (-\lambda)$, the later
corresponding to the interchange
of $\mu\leftrightarrow (1-\mu)$.

We first consider the quasistatic limit $\epsilon_a\ll \epsilon_b$,
viz $a(t)\simeq a$ is approximatively constant.
The fixpoint $(y^*,b^*)$ in the $(y,b)$ plane
is then determined by
\begin{equation}
\begin{array}{rcl}
0 &=& \lambda \left(y^*\right)^2-(2+\lambda)y^*+1 \\
b^* &=& g^{-1}(y^*)-a[y^*-1/2] \\
    &=& \log(y^*/(1-y^*))-a[y^*-1/2]
\end{array}
\label{eq_fixpoint_yb}
\end{equation}

A straightforward linear stability analysis shows, 
that the fixpoint $(y^*,b^*)$ remains stable for small
gains $a$ and becomes unstable for large gains, see
Fig.\ \ref{fig_stability}. We now go beyond the
quasistatic approximation and consider a small but
finite polyhomeostatic adaption rate $\epsilon_a$ for the gain.
Starting with a small gain $a$ we see, compare
Eq.\ (\ref{eq_one_site}), that the gain necessarily grows
until it hits the boundary towards instability; for
small $\Delta\tilde b$ the growths of the gain
is $a(t)\sim \sqrt{t}$.

\begin{figure}[t]
\centerline{
\includegraphics[width=0.40\textwidth]{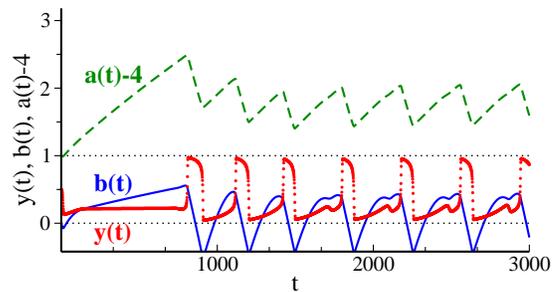}
           }
\caption{(color online) The time dependence of $y(t)$ (red stars)
$b(t)$ (solid line) and of $a(t)-4$ (dashed line) for
the balanced one-site problem (\ref{eq_one_site}).
$\epsilon_a=\epsilon_b=\epsilon=0.01$ and
$\mu=0.28$; the horizontal dotted lines are guides to the eye.
The gain $a(t)$ is initially small and the system
relaxes, since $\epsilon\ll1$, fast to a fixpoint 
of $y=g(a[y-1/2]+b)$. Once $a(t)$ surpasses a certain threshold,
compare Fig.\ \ref{fig_stability}, the fixpoint 
becomes unstable and the system starts to spike spontaneously.
}
\label{fig_oneSiteTime} 
\end{figure}

In other words, a finite adaption rate $\epsilon_a$ 
for the gain turns the fixpoint
attractor $(y^*,b^*)$ into an attractor relic and the
resulting dynamics becomes non trivial and autonomously
self-sustained. This route towards autonomous neural 
dynamics is actually an instance of a more general
principle. Starting with an attractor dynamical system,
viz with a system governed by a finite number of stable
fixpoint, one can quite generally turn these fixpoints 
into unstable attractor ruins by coupling locally 
to slow degree of freedoms~\cite{gros07,gros09}, in
our case the slow local variables are $a(t)$ and $b(t)$. 
Interestingly, there are no saddlepoints present 
in attractor relic networks in general,
and in (\ref{eq_one_site}) in particular, and
hence no unstable heteroclines, as in a proposed
alternative route to transient state dynamics
via heteroclinic switching.~\cite{rabinovich01,kirst08} 

In Fig.\ \ref{fig_oneSiteTime} the time evolution
is illustrated for $\lambda=3.017$, which corresponds to 
$\mu=0.28$, see Eq.\ (\ref{eq_exponential}),
and $\epsilon_a=\epsilon_b=\epsilon=0.01$.
The system remains
in the quasistationary initial regime until the
gain $a$ surpasses a certain threshold. The initial
quasistationary fixpoint becomes therefore unstable
via the same mechanism discussed analytically above,
see Eq.\ (\ref{eq_fixpoint_yb}),
for the regime $\epsilon_a\to 0$, compare
Fig.\ \ref{fig_stability}. The output activity
$y(t)$ oscillates fast between two transient
fixpoints of $y=g(a[y-1/2]+b)$, having a high and
a low value respectively.
This spiking behavior of the neural 
activity is driven by spontaneous 
oscillations in the threshold 
$-b(t)/a(t)$, shifting the intersection of
$g(a[y-1/2]+b)$ with $y$ forth and back. 

Evaluating the local Lyapunov exponents we find that
the trajectory is stable against perturbations 
for the transient states close to one of the 
transient fixpoints of $y=g(a[y-1/2]+b)$ and
sensitive to perturbations and external modulation
during the fast transition periods, an instantiation
of the general notion of transient state
dynamics~\cite{gros07,gros09}.

\begin{figure}[t]
\centerline{
\includegraphics[width=0.40\textwidth]{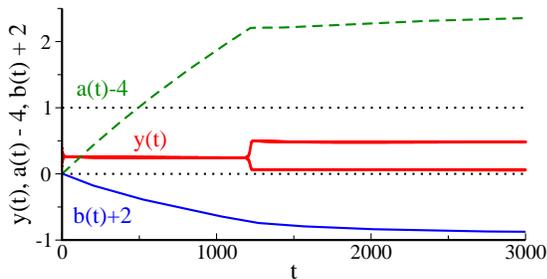}
}
\caption{(color online) The time dependence of 
$y(t)$ (red stars) $b(t)+2$ (solid line) and of 
$a(t)-4$ (dashed line) for the one-site problem
with inhibitory self-coupling $x\to1/2-y$,
$\epsilon_a=\epsilon_b=\epsilon=0.01$ and
$\mu = 0.28$. A Hopf-bifurcation occurs for the
output $y(t)$ when the initial quasistationary fixpoint
becomes unstable, giving place to a new fixpoint
of period two.
}
\label{fig_inverse} 
\end{figure}


{\it Inhibitory self-coupling.}--
So far we discussed, see 
(\ref{eq_one_site}) and Fig.\ \ref{fig_oneSiteTime},
a neuron having its output $y(t)$ coupled back 
excitatorily to its input via $x\to y-1/2$.
The dynamics changes qualitatively for an
inhibitory self-coupling $x\to 1/2-y$, see
Fig.\ \ref{fig_inverse}. There is now only
a single intersection of $g(a[-y+1/2]+b)$ with $y$.
This intersection corresponds to a stable fixpoint for small
gains $a$. A Hopf-bifurcation~\cite{gros08} 
occurs when
$a(t)$ exceeds a certain threshold and a new
fixpoint of period two becomes stable. The
coordinates of this fixpoint of period two
slowly drift, compare Fig.\ \ref{fig_inverse},
due to the residual changes in $a(t)$ and
$b(t)$. Interestingly, the dynamics remains
non-trivial, as a consequence of the continuous
adaption, even in the case of inhibitory
self-coupling.


{\it Self organized chaos.}--
We have studied numerically fully connected 
networks of $i=1,..,N$ polyhomeostatically 
adapting neurons (\ref{eq_intrinsic_plasticity}), 
coupled via
\begin{equation}
x_i(t) \ =\ {\textstyle \sum_{j\ne i}} w_{ij} y_j(t)~.
\label{eq_couplings}
\end{equation}
The synaptic strengths are $w_{ij}=\pm 1/\sqrt{N-1}$,
with inhibition/excitation drawn randomly with
equal probability. The adaption rates are
$\epsilon_a=\epsilon_b=0.01$. We consider
homogeneous networks where all neurons
have the identical $\mu$ for the target output 
distributions (\ref{eq_exponential}),
with $\mu=0.15,\, 0.28$ and $\mu=0.5$.

\begin{figure}[t]
\centerline{
\includegraphics[width=0.40\textwidth]{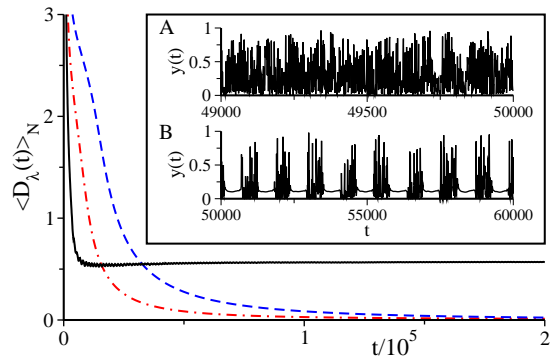}
}
\caption{(color online) Mean Kullback-Leibler
divergence $D_\lambda$, for $N=500$ networks,
of the real and target output distributions as 
a function of number of iterations $t$. The solid black, 
dashed-dotted red and the dashed blue lines
correspond respectively to target firing rates
$\mu=0.15$, $\mu=0.28$, and $\mu=0.5$.
Inset: Output activity of two randomly chosen 
neurons for target average firing rates $\mu=0.28$ (top) 
and $\mu=0.15$ (bottom).
       }
\label{fig_outputActi} 
\end{figure}

The activity patterns presented in the inset of 
Fig.\ \ref{fig_outputActi} are chaotic 
for the $\mu=0.28$ network and laminar
with intermittent bursting for the network
with $\mu=0.15$. This behavior is typical
for a wide range of network geometries,
distributions of the synaptic weights and
for all initial conditions sampled. The
respective Kullback-Leibler divergences $D_\lambda$
decrease with time passing, as shown in
Fig.\ \ref{fig_outputActi}, due to the
ongoing adaption process (\ref{eq_intrinsic_plasticity}).
$D_\lambda$ becomes very small, though still finite,
for long times in the self-organized chaotic regime 
($\mu=0.28,\, 0.5$), remaining
substantially larger in the intermittent
bursting phase ($\mu=0.15$).

We have evaluated the
global Lyapunov exponent,~\cite{ding07} 
finding that two initial close orbits diverge 
until their distance in phase space corresponds,
within the available phase space, to the distance of 
two randomly selected states, the tellmark sign of 
deterministic chaos.~\cite{gros08} The corresponding
global Lyapunov exponent is about 5\% per time-step 
for $\mu=0.5$ and initially increases with the decrease 
of $\mu$, until the intermittent-bursting regime emerges, 
after which the global Lyapunov exponent declines
with the decrease of $\mu$.

The system enters the chaotic regime, in close
analogy to the one-side problem discussed previously,
whenever the adaptive dynamics 
(\ref{eq_intrinsic_plasticity}) has pushed the 
individual gains $a_i(t)$, of a sufficient number 
of neurons, above their respective 
critical values. Hence, the chaotic state
is self-organized. Chaotic dynamics
is also observed in the non-adapting limit, with
$\epsilon_a,\epsilon_b\to0$, whenever the
static values of $a_i$ are above the critical
value,
in agreement with the results of a large-$N$ 
mean field analysis of an analogous continuous 
time Hopfield network.~\cite{sompolinsky88}
Subcritical static $a_i$ lead on the other side
to regular dynamics controlled by point attractors.

\begin{figure}[t]
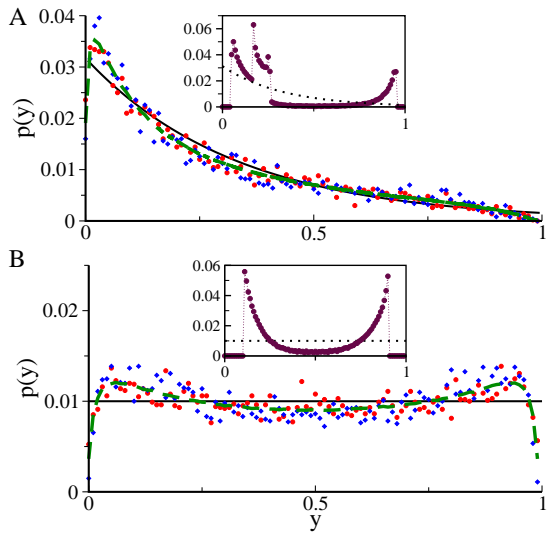

\begin{center}$
\begin{array}{c}
\includegraphics[width=0.4\textwidth]{outDistA.eps} \\
\includegraphics[width=0.4\textwidth]{outDistB.eps}
\end{array}$
\end{center}

\caption{(color online) Output distributions of the two neurons
with highest (blue diamonds) and lowest (red circles) 
Kullback-Leibler divergence (\ref{eq_Kullback_Leibler})
compared to the mean output distribution 
(dashed green line) and the target exponential 
output distribution (full black line). For a
network with $N=500$ neurons and a target output 
mean firing rate (A) $\mu = 0.28$ and (B) $\mu = 0.5$ 
(identical for all neurons).
Insets: For the same parameters the mean output 
distribution of the single self-coupled neuron,
see Fig.~\ref{fig_stability}.
}
\label{fig_outdist} 
\end{figure}

In Fig.\ \ref{fig_outdist} we present the
distribution of the output activities for networks
with target mean firing rates $\mu=0.28$ 
and $\mu=0.5$. Shown are in both cases the 
firing rate distributions
$p(y)$ of the two neurons having 
respectively the largest and the smallest
Kullback-Leibler divergence (\ref{eq_Kullback_Leibler})
with respect to the target
exponential distributions (\ref{eq_exponential}).
Also shown in Fig.\ \ref{fig_outdist} are the 
distributions of the network-averaged output activities.

The data presented in Fig.\ \ref{fig_outdist} 
shows, that the polyhomeostatically self-organized 
state of the network results in firing-rate 
distributions close to the target distribution.
This result is quite remarkable. The polyhomeostatic
adaption rules (\ref{eq_intrinsic_plasticity})
are local, viz every neuron adapts its internal
parameters $a_i$ and $b_i$ independently 
on its own. 

{\it Intermittency.}--
Interestingly, the neural activity presented 
in the inset of Fig.\ \ref{fig_outputActi} for 
$\mu=0.15$ shows intermittent or bursting behavior.
In the quiet laminar periods the neural activity 
$y(t)$ is close to (but slightly below) the target 
mean of $0.15$, as one would expect for a 
homeostatically regulated system and the local 
Lyapunov exponent is negative. The target
distribution of firing rates (\ref{eq_exponential})
contains however also a small but finite weight
for large values for the output $y(t)$, which
the system achieves through intermittent bursts
of activity. In the laminar/bursting periods 
the gain $a(t)$ is below/above threshold, acting 
such as a control parameter~\cite{platt93}. 

Both the intermittent and the chaotic regime are
chaotic in terms of the global Lyapunov being 
positive in both regimes. A qualitative difference
can be observed when considering the subspace
of activities $(y_1,\dots,y_N)$. Evaluating
the global Lyapunov exponent in this subspace, viz
the relative time evolution 
$(\Delta y_1,\dots,\Delta y_N)$ of
differences in activities, we find this restricted
global Lyapunov to be negative in the intermittent
and positive in the chaotic regime. Details on the
evolution from intermittency to chaos 
as a function of $\mu$ and system sizes
will be given elsewhere.


{\it Conclusions.}--It is well known~\cite{arbib02}, 
that individual and networks of spiking neurons may 
show bursting behavior. Here we showed, that networks 
of rate encoding neurons are intermittently bursting 
for low average firing rates, when polyhomeostatically 
adapting, entering a fully chaotic regime for larger 
average activity level. Autonomous and polyhomeostatic
self-control of dynamical systems may hence lead quite
in general to non-trivial dynamical states and novel phenomena.

Polyhomeostatic optimization is of relevance in a range
of fields. Here we have discussed it in the framework
of dynamical system theory. Alternative examples
are resource allocation problems, for which a given
limited resource, like time, needs to be allocated to
a set of uses, with the target distribution being the
percentages of resource allocated to the respective uses.
These ramifications of polyhomeostatic optimization will 
be discussed in further studies.



\section*{References}
\begin{thebibliography}{99}

\bibitem{turrigiano99} G.G. Turrigiano, 
Trends Neurosci {\bf 22}, 221 (1999).
%
\bibitem{davis06} G.W. Davis, 
Annual review of neuroscience {\bf 29}, 307 (2006).
%
\bibitem{marder06} E. Marder, and J.M. Goaillard, 
Nature Reviews Neuroscience {\bf 7}, 563 (2006).
%
\bibitem{der99} R. Der, U. Steinmetz, and F. Pasemann, 
in {\sl Computational Intelligence for Modelling, Control, and
Automation, Concurrent Systems Engineering} {\bf 55}, 43 (1999). 

\bibitem{stemmler99} M. Stemmler, C. Koch.
Nature Neuroscience {\bf 2}, 521 (1999).
%
\bibitem{triesch05a} J. Triesch,
in {\sl Proceedings of ICANN 2005}, W. Duch  et al.\ (Eds.), 
LNCS {\bf 3696}, 65 (2005).
%
\bibitem{cover06} T.M. Cover, J.A. Thomas, 
{\it Elements of information theory},
Wiley 2006.
%
\bibitem{seung00} H.S. Seung, D.D. Lee, B.Y. Reis, and D.W. Tank, 
J. Comp. Neuroscience {\bf 9}, 171 (2000).
%
\bibitem{steil07} J.H. Steil, 
Neural Networks {\bf 20}, 353 (2007).
%
\bibitem{gros07} C. Gros,
New Journal of Physics {\bf 9}, 109 (2007).
%
\bibitem{gros09} C. Gros,
Cognitive Computation {\bf 1}, 77 (2009).
%
\bibitem{rabinovich01} M. Rabinovich {\it et al},
Phys. Rev. Lett. {\bf 87}, 68102 (2001).

\bibitem{kirst08} C. Kirst, and M. Timme,
Phys. Rev. E {\bf 78}, 065201 (2008).

\bibitem{gros08} C. Gros,
{\it Complex and Adaptive Dynamical Systems, A Primer},
              Springer (2008).


\bibitem{ding07} R. Ding, and J. Li, 
Phys. Lett. A {\bf 364}, 396 (2007).

\bibitem{sompolinsky88} H. Sompolinsky, A. Crisanti, and H.H. Sommers, 
Phys. Rev. Lett. {\bf 61} 259 (1988).

\bibitem{arbib02} M.A. Arbib, 
{\sl The handbook of brain theory and neural networks},
  MIT Press (2002).

\bibitem{platt93} N. Platt, E.A. Spiegel, and C. Tresser, 
Phys. Rev. Lett. {\bf 70}, 279 (1993).

\end{thebibliography}
\end{document}